\documentclass[a4paper,11pt]{article}
\usepackage{pos}
\usepackage{subcaption}
\usepackage{csquotes}

\title{International Lattice Data Grid 2.0: Status and Progress}

\author[a]{Basavaraja Bheemalingappa Sagar}
\affiliation[a]{CTAO, Science Data Management Centre (SDMC), Platanenallee 6,
  15738 Zeuthen, Germany}
\emailAdd{basavaraja.bheemalingappa.sagar@desy.de}

\author[b]{Georg von Hippel}
\affiliation[b]{PRISMA$^+$ Cluster of Excellence and Institut f\"ur Kernphysik,
  Johannes Gutenberg-Universit\"at Mainz, Johann-Joachim-Becher-Weg 45, 55099 Mainz, Germany}
\emailAdd{hippel@uni-mainz.de}

\author[c]{Giannis Koutsou}
\affiliation[c]{Computation-based Science and Technology Research Center,
  The Cyprus Institute, 20 Kavafi Str., Nicosia 2121, Cyprus}
\emailAdd{g.koutsou@cyi.ac.cy}

\author[d]{Hideo Matsufuru}
\affiliation[d]{Computing Research Center, High Energy Accelerator
  Research Organization (KEK), and Accelerator Science Program,
  Graduate Institute for Advanced Studies, Graduate University for
  Advanced Studies (SOKENDAI), Oho 1-1, Tsukuba 305-0801, Japan}
\emailAdd{hideo.matsufuru@kek.jp}

\author[e]{Dirk Pleiter}
\affiliation[e]{Bernoulli Institute for Mathematics, Computer Science and Artificial Intelligence,
  University of Groningen, 9700 AK Groningen, The Netherlands}
\emailAdd{d.h.pleiter@rug.nl}

\author[f]{Hubert Simma}
\affiliation[f]{John von Neumann-Institut für Computing NIC, Deutsches Elektronen-Synchrotron DESY,
  Platanenallee 6, 15738 Zeuthen, Germany}
\emailAdd{hubert.simma@desy.de}

\author*[g]{Carsten Urbach}
\affiliation[g]{Helmholtz-Institut f\"ur Strahlen- und Kernphysik (Theorie) and Bethe,
  Center for Theoretical Physics, Universit\"at Bonn, 53115 Bonn, Germany}
\emailAdd{urbach@hiskp.uni-bonn.de}

\abstract{
  In this proceeding contribution we discuss the status and progress
  towards a modernised and extended  International Lattice Data Grid
  (ILDG), which has seen major 
  developments, updates, and upgrades over the last year. In
  particular, metadata and file schemata have been extended. Moreover,
  the registration and authentication services have been modernised,
  and the file and metadata catalogues re-implemented.
}

\FullConference{The 41st International Symposium on Lattice Field Theory (LATTICE2024)\\
 28 July - 3 August 2024\\
Liverpool, UK\\}

\begin{document}
\maketitle

\section{Introduction}

The International Lattice Data Grid (ILDG)~\cite{Davies:2002mu,Irving:2003uk,Ukawa:2004he,Beckett:2009cb,ildg-organization}
represents a joint effort 
of the Lattice QCD (LQCD) community, which started some 20 years ago
with the goal of sharing valuable gauge configurations worldwide. The
ILDG was implemented as a federation of autonomous regional grids,
CSSM~\cite{cssm-organization} for Australia, JLDG for Japan~\cite{jldg-organization}, Latfor DataGrid
(LDG)~\cite{ldg-organization} for 
continental Europe, UK Lattice Field Theory for the
UK~\cite{uklft-organization}, and 
USQCD~\cite{usqcd-organization} for 
the US.

Each regional grid autonomously operates the ILDG core services: a
website, a metadata catalogue (MDC), a file catalogue (FC), and one or more
storage elements (SE). At the global level, ILDG provides a set of federation services.
This, in particular, includes a user registration and authentication
service.
ILDG users are organised as a virtual organisation (VO).
For managing this VO, originally the virtual organisation membership service (VOMS) was used.

The ILDG as an organisation is borne by the ILDG board, the Metadata
Working Group (MDWG), and the Middleware Working Group (MWWG). The ILDG
board is responsible for administrative and organisational matters of
the ILDG. The MDWG is responsible for the specification and update of
metadata schemata and file formats, and the MWWG for the underlying
services.

While not explicitly intended at the time, the ILDG
implements the FAIR principles, i.e Findability, Accessibility,
Interoperability, and Reusability, which were only published in
2016~\cite{wilkinson2016fair}. 

Due to a lack of resources, the ILDG services unfortunately degraded
over the last decade. Only due to funding obtained as part of the NFDI
(national research data infrastructure~\cite{nfdi-organization}) funding
scheme of the German 
Research Foundation (DFG) and the NFDI-funded PUNCH4NFDI (Particles,
Universe, Nuclei and Hadrons for the NFDI)
consortium~\cite{punch4nfdi}, it became 
possible to refurbish, modernise, and resume the ILDG services. In
parallel, a significant effort was made to update and extend the metadata
schemata and file formats to serve today's requirements of the LQCD
community and beyond.

These efforts led to \enquote{ILDG 2.0}, a modernised version of the ILDG~\cite{Karsch:2022tqw}.
This includes a new user registration and authentication service to replace VOMS.
Now a dedicated INDIGO IAM~\cite{indigo-iam} instance hosted by
INFN-CNAF is used.
A key benefit is the support of token-based authentication, which is widely used for today's cloud infrastructures.
Furthermore, the metadata and
file catalogue services have been reimplemented, which can be accessed through REST APIs.
Deployment of these services has been improved by means of containers.
New versions of the configuration and ensemble metadata schemata,
as well as the specification of an extended file format are ready
for the markup and sharing of the gauge configurations. Prototype
implementations of simple client tools with command-line, graphical,
or web interfaces are available.
Details will be discussed in the remainder of this proceedings contribution.

\section{Metadata Schemata}

Metadata schemata aim to provide means to markup data in a
standardised and unique way. At the same time, such schemata should be
easily extensible and flexible to handle. The Extensible Markup
Language (XML) fulfils these requirements: it is a W3C
standard~\cite{w3c}, it 
provides powerful query and search technology via XPath, and
well-tested and open-source software implementations are widely
available. Moreover, with the XML Schema Definition (XSD), schemata
and unique data markup can be standardised. While mainly aimed
at machine readability, XML is also human-readable to a certain extent.
For these reasons, the ILDG schemata are based on XSD, which worked
very successfully since the beginning.
Nowadays, JSON or YAML could be used as alternatives to XML. However, for
backward compatibility, ILDG will stick to XML/XSD. Also, there are no
obvious advantages of JSON or YAML over XML.

The structure of the markup in ILDG is as follows:
The binary data (gauge configurations) are  stored on a storage element and marked up
by a separate \emph{configuration XML document}. This configuration XML
document is required to validate against the ILDG schema, called QCDmlConfig. Apart
from some configuration-specific information, like for instance the
plaquette value and a checksum, the configuration XML file links to the binary file
via the logical file name (LFN), but also to the Markov chain or ensemble the
configuration belongs to via the Markov Chain Uniform Resource
Identifier (markovChainURI). Finally, the ensemble is marked up using
an \emph{ensemble XML document}, which needs to validate against the so-called
QCDmlEnsemble schema of ILDG. This XML document compiles all information common to all
configurations that are part of one ensemble. This information
includes for instance the algorithm description and the lattice action.

For the extension of the file format~\cite{file-format}, and of the
QCDml schemata~\cite{qcdml-git}, the community was queried, and the
MDWG reviewed, discussed and 
implemented most of the requests\footnote{A detailed description of
the changes and extensions can be found on the web-page~\cite{ildg-specs}}.

To provide some means for provenance tracking, ILDG does foresee entries in the XML files,
where certain actions (e.g., generation of gauge configurations) can be attributed to so-called participants.
In the updated schemata, it is now possible to 
specify the ORCID as an alternative to
specifying the name and institution in the \texttt{<participant>}
element. The participant can be identified by an \texttt{<orcid>}
element, followed by optional \texttt{<name>} and/or
\texttt{<institution>}, or by the latter two only.

Moreover, at top-level there is a new
\texttt{<additionalInfo>} element (and optional sub-tree), which can be
used to provide additional information. Several other elements allow
now for a new \texttt{<annotation>} sub-element in order to be able to
provide more specific additional information. The new
\texttt{<annotation>} element also replaces the former
\texttt{<comment>} element.

\subsection{Ensemble XML}

The ensemble XML schema was extended by three major changes. First,
a license must be specified as part of the ensemble XML, which
is a requirement to be compliant with the FAIR principles. One can
specify standard licenses, like e.g. a Creative Commons public license, or a custom
license. This \texttt{<license>} element comes with two important
additional possibilities: one is the possibility to specify an embargo
date, the other is to provide a way to specify how the usage of the
ensemble must be acknowledged in a respective publication.

Second,
there is now the possibility to specify a funding reference via one or
more entries in a non-empty list of
\texttt{<fundingReference>} elements. The MDWG decided to follow
here the example of the DataCite schema~\cite{datacite}, however, restricted to a
subset of it. Every funding reference has a mandatory
\texttt{<funderName>} element and optional \texttt{<awardTitle>}
and/or \texttt{<awardNumber>} child elements.

Third, the list of lattice actions and gauge groups has been
significantly extended to reflect the progress in the field. The
supported gauge groups are now SU$(N)$ and SO$(N)$ for $N\geq 2$,
SP$(N)$ with even $N\geq 4$ and U$(N)$ with $N\geq1$.

On the lattice action side, several new action types have been
defined. To account for the growing number of works including QED
effects, QED gauge actions are available as new XML elements.
Higher fermion representations are now
possible to markup, including for instance fermions in the adjoint
representation. Open boundary conditions with and without Schrödinger
functional can be marked up, Möbius domain-wall fermions as well as
Wilson clover fermions with exponential clover term.

Finally, it is possible to specify fermions coupled to electromagnetic
fields. Initially, the ensemble XML schema includes so-called SLINC
fermions as used by QCDSF and the coupling in the framework of
$C^\star$ boundary conditions. In compact form, all the added action
specifiers are the following: 

\noindent{\footnotesize
\begin{tabular}{ll}
\texttt{<wilsonAdjointQuarkAction>} &
\texttt{<wilsonTwoIndexSymmetricQuarkAction>} \\
\texttt{<wilsonTwoIndexAntisymmetricQuarkAction>} &
\texttt{<treelevelSymanzikOpenBCGluonAction>}\\
\texttt{<npCloverOpenBCQuarkAction>} &
\texttt{<moebiusDomainWallQuarkAction>}\\
\texttt{<nonCompactQedSPhotonAction>} &
\texttt{<compactPlaquetteCstarPhotonAction>}\\
\texttt{<compactTreelevelSymanzikCstarPhotonAction>}&
\texttt{<fatLinkDerivNpNiCloverChargedQuarkAction>}\\
\texttt{<npCloverCstarChargedQuarkAction>}&
\texttt{<npExpCloverQuarkAction>}\\
\end{tabular}
}\\


\subsection{Configuration XML}

The main change in the ILDG configuration XML schema is due to the
request to store multiple gauge configurations in a single binary
file. In order to make this possible, the one-to-one correspondence
between one single gauge configuration binary data file and one single
configuration XML file had to be abandoned. Instead of a single
\texttt{<markovStep>} element, there is 
now a single \texttt{<markovSequence>} element per configuration XML
file. The \texttt{<markovSequence>} can have one or more
\texttt{<markovStep>} elements as child elements. The latter have the
checksum stored in \texttt{<crcCheckSum>} and the plaquette in
\texttt{<avePlaquette>} sub-elements. For consistency, the Markov
Chain URI (\texttt{<markovChainURI>}) and the \texttt{<series>}
elements are now mandatory sub-element of \texttt{<markovSequence>}. 


\subsection{ILDG file format}

The information contained in ILDG data files has three components:
the link variables of a gauge configuration in binary format, minimal format
and layout information to unambiguously extract the link variables, and the unique
identifier (the site-independent LFN) of the data and the associated metadata.
The ILDG file format specifies the content of these three
components and the way how they are packed according to the LIME
format~\cite{lime} into a single, structured file. The previous
version 1.1 of 
the file format specification only allowed the packing of a single SU$(3)$
gauge configuration.

The new version 1.2 of the file format adds extensions to the previous
specifications, such that existing ILDG binary files remain valid.
In particular, the new version supports packing of multiple gauge
configurations, gauge groups different from SU$(3)$ (e.g. for
QCD+QED or Beyond-Standard-Model simulations), and reduced
storage formats (e.g. storing only the first $N-1$ rows of SU$(N)$
link variables).

\section{ILDG Core Services}

Having a unified user registration and a single-sign-on (SSO) service is a key element
for the ILDG data infrastructure, in order to enable uniform, safe, and controlled data access.
The user registration and SSO are --- apart from the ILDG web page --- the
only service that is operated under the joint and direct responsibility of ILDG as a
whole. In contrast, the actual storage resources and catalogue services are 
set up and operated autonomously by the individual regional grids, i.e. according
to their specific technical choices, organisational setup, and funding possibilities.

Interoperability between the services and infrastructure of the regional grids
is achieved by following the common ILDG specifications. Most importantly, they define
--- in addition to the metadata schema and file format, as discussed above ---
a common API for the catalogue services.

A main goal of the technical modernisation of ILDG 2.0 was the migration
from grid certificates to a token-based authentication and authorisation setup.
This required major changes to all components of ILDG, in particular, setting
up of a new user registration and SSO service, a complete re-factoring of
the catalogue services, and the re-configuration of all storage elements.

Initially, it was foreseen to have an intermediate phase in which both,
grid certificates and tokens, are used in parallel, like in WLCG. However,
in early 2024 it became clear that a more drastic and ambitious transition
to a completely token-based setup might be simpler and more
effective.

\subsection{User Registration}

User registration and authentication is an established requirement within the LQCD community
(see, e.g., \cite{Bennett:2025xai}).
This is for different reasons.
First of
all, requesting users to become members of the virtual 
organisation ILDG helps to ensure that they know and accept the rules
and conditions that govern the use of resources and services (i.e.
the relation between users and resource providers), as well as the
sharing of gauge configurations between users (acting as data
providers and data consumers).

Second, the storage providers, e.g. HPC centres, have to provide in
addition to the storage space also network bandwidth enabling
access to the data. This infrastructure is expensive and must not be
saturated with arbitrary downloads, e.g. by bots or anonymous attackers.
Therefore, even for read-only access to large data sets, most service
providers require authentication of the users with reliable identity
vetting. Of course, also uploading or modifying data on the storage
elements should always be protected by some level of access control
and authorisation.

Finally, many members of the LQCD community also request the possibility
of having so-called embargo periods, which are a way to allow
collaborations who generated the data to be the first exploiters of the data. The main motivation
for this request is the fact that the production of gauge configuration
ensembles represents a significant effort.
On the other hand, timely upload of configurations to ILDG right after their production
and already during the embargo period is advisable to avoid unnecessary duplication
of storage space or extra efforts for data movement and maybe markup at a later date.
Furthermore, leveraging the ILDG data infrastructure can be highly beneficial for
sharing data within the data-producing collaborations and splitting the work related to the processing of this data
(e.g. computation of physical observables on top of these configurations).
This requires the availability of simple and reliable access control mechanisms
to temporarily restrict read access to specific users or collaborations.

In ILDG 2.0, a new user registration service was deployed based on an INDIGO Identity and Access Management (IAM) service.
Users can now request ILDG membership at
this service after authentication through a trusted Identity Provider (IdP), which in most cases can be their home institution.
Many of these institutions are already part of the international inter-federation service eduGAIN.
For this purpose, the IdP must release the required user attributes. This is
usually the case for institutions that are part of eduGAIN (for details, see~\cite{EARC}).

The INDIGO IAM software~\cite{indigo-iam} has been developed by INFN-CNAF and is used by several
scientific communities, e.g. LHC experiments or SKA. In addition to user authentication
and group management, the IAM issues the OIDC/Oauth2 tokens needed to access
other ILDG services and resources. Thanks to a built-in scope-policy engine and
many further configuration options, the IAM can thus also act as an authorisation service.

A dedicated INDIGO IAM instance for ILDG~\cite{iam-ildg} is hosted and operated by INFN-CNAF in Bologna.
Being an essential component of ILDG, various non-technical aspects had to be taken
care of, including a memorandum of understanding and new policy documents (data
sharing policy, acceptable use policy, and privacy notice).

\subsection{Metadata and File Catalogue}

The possibility to register and search metadata is an essential
requirement of the FAIR principles~\cite{wilkinson2016fair}. In
ILDG the descriptive metadata, i.e. the ensemble and configuration XML
documents, are kept in the metadata catalogues (MDC).
Resolving the unique identifiers recorded in the metadata documents to possibly multiple
storage locations, which may change over time, happens through file catalogue (FC) services.

For ILDG 2.0, a new FC service had to be implemented and the legacy MDC from LDG
was completely re-factored. Both catalogues have now a REST API, which
implements the new ILDG specifications, and are deployed as containers.
The common API guarantees interoperability between regional grids or with
additional services (see also~\cite{Pederiva}), while the containerisation
allows simple deployment of multiple instances of the same catalogue
in the different regional grids to save development effort.

Metadata in the new MDC can be searched either by XPath queries
(typically used for ensemble XML documents) or by a configurable
set of \enquote{quick-search keys} (e.g. the \texttt{markovChainURI}
of configuration XML documents).
The XPath queries allow for very different kinds of searches but may be (very) slow.
The quick-search keys provide a fast but restricted search mechanism.
   
The MDC is capable of holding different types of XML document collections, each with a
separate metadata schema (specified by an XSD document) and a number
of configurable properties, like different mechanisms and levels of
controlling read and write access, and the set of supported quick-search
keys (defined by XSLT transformations). This capability can in the future be exploited to
extend ILDG to sharing data objects beyond gauge configurations, e.g. correlation
functions, analysis workflows, published ensembles, or from other
(non-Lattice) domains.

Token-based authorisation for the access to (meta-) data objects
in the MDC or FC is implemented in an analogous way as in the storage
elements (see below): authorisation is decided on the basis of a suitable
comparison between the path information in the \enquote{\texttt{scope}} claim
of the access token and some path-like attribute of the protected data object.


\subsection{Storage Elements}

ILDG 2.0 follows the WLCG profile~\cite{wlcg-profile} to implement fine-grained
capability-based authorisation via Oauth2 access tokens. Such an access control
could possibly have a granularity as fine as referring to individual files.
For instance, a user who has a valid access token with scope \texttt{storage.read:/x/y}
(together with the appropriate issuer and audience claims) can only read files which are
in the directory sub-tree with the root at the path \texttt{x/y} relative to the base URL of
the storage element.

The storage elements currently available in LDG use either dCache or StoRM
middleware, which can be configured to work with capability-based authorisation
via Oauth2 tokens. Therefore, all four storage elements of LDG are meanwhile fully
enabled for token-based access.
In JLDG the development work for enabling access via tokens to the Gfarm storage is in progress.

In general, any disk- or tape-storage server can act as a storage element within
ILDG provided that it has an interface which (i) supports an access protocol foreseen within ILDG,
e.g. webDaV/http, and (ii) recognises and respects the authorisation information
carried by the access token which the user presents when requesting access to data.
In order to make new storage elements available to ILDG, it is necessary to  
also explore cloud storage technologies and lightweight storage interfaces,
which can easily be set up by smaller groups or institutions to expose their local
storage resources to ILDG.

\section{Conclusions and Outlook}

Major progress has been made towards ILDG 2.0 as a modernised and extended version
of the ILDG, including revised metadata schemata, a new ILDG-wide user registration,
re-factored catalogue services, and token-based (meta-)data access.

Several related circumstances seem to have been important for this progress: (i) the
re-activation of the ILDG Board and of the Metadata and Middleware Working Groups,
bringing together persons who push forward the solution of technical and organisational
challenges of ILDG in their regional grids and local institutions; (ii) the
availability of a professional software developer thanks to funding within the
PUNCH4NFDI project; and (iii) the increased importance that is given to data sharing
in science and the open science vision.

Active support and participation of the LQCD community is certainly necessary
to further improve and sustain ILDG in the future. In particular, storage resources
or person-power for the development and maintenance of ILDG services need to be
considered worthwhile to be included in funding applications.

The next steps towards ILDG 2.0 concern the re-activation
of still missing services in the various regional grids, consistently
restoring the metadata of the legacy ensembles which are still available
from ILDG 1.0, and the start of massive uploads of new data. To make these possible
and more practicable for a wider user community, further improvement, in
particular of client tools and documentation, is definitely needed. Also,
a community-wide hands-on workshop for training is desirable in the near
future. Further minor revisions of the metadata schema, if needed, should
be expected to be released timely, and a further revision of the ILDG file
format could include support of HDF5 (instead of LIME) as a packing format.

We would encourage those interested in driving such work forward
to contact us,
and to join relevant mailing lists of the ILDG~\cite{ildg-organization}.


\acknowledgments
The progress towards ILDG 2.0 is due to the effort and help of many people.
We thank all members of the ILDG Board and of the Metadata and Middleware
Working Groups for their efforts and support.
We are grateful to Y.~Nakamura, C.~McNeile, H.~Ohno, and C. Schmidt for
deployment tests of the catalogues, to A.~Patella for his valuable contributions
and advice towards the new revision of QCDml, and to many colleagues from the
PUNCH4NFDI consortium for helpful discussions and contributions.

We thank the INDIGO IAM team at INFN-CNAF, in particular, F. Agostini, F. Giacomini, R. Miccoli,
C. Pellegrino, A. Rendina, and E. Vianello for developing and hosting the IAM service for ILDG, and for their valuable help and
advice. We also wish to thank the various IT teams, e.g. at CNAF (Bologna), DESY (Hamburg and Zeuthen),
Forschungszentrum Jülich, Tsukuba University (Japan), and many others, for providing important
services and support.

The work of B.~B.S. was part of the PUNCH4NFDI project and funded by the Deutsche
Forschungsgemeinschaft (DFG, German Research Foundation) --- project number 460248186.

G.~K. acknowledges support from EXCELLENCE/0421/0195, co-financed by the European Regional
Development Fund and the Republic of Cyprus through the Research and Innovation Foundation and the
AQTIVATE Marie Sk\l{}odowska-Curie Doctoral Network GA~No.~101072344.

The work of C.U. on this project was supported by the
Deutsche Forschungsgemeinschaft (DFG, German Research Foundation) as
part of the CRC 1639 NuMeriQS – project no. 511713970. 

\paragraph*{Open access statement}
For the purpose of open access, the authors have applied a Creative Commons
Attribution (CC BY) licence to any Author Accepted Manuscript version arising.

\paragraph*{Research Data Access Statement}
No new data were generated during the preparation of this work.


\bibliography{references}
\bibliographystyle{JHEP}
\end{document}